# Single Crystalline Silver Films for Plasmonics: From Monolayer to Optically Thick Film

*Fei Cheng,[†] Chien-Ju Lee,[∥] Junho Choi,[†] Chun-Yuan Wang,[†,‡] Qiang Zhang,[†] Hui Zhang,[†] Shangjr Gwo,[‡] Wen-Hao Chang,[∥] Xiaoqin Li[†] and Chih-Kang Shih*,[†]*

[†]Department of Physics, University of Texas at Austin, Texas 78712 United States

[‡]Department of Physics, National Tsing-Hua University, Hsinchu 30013, Taiwan

[∥]Department of Electrophysics, National Chiao Tung University, Hsinchu 30010, Taiwan

**ABSTRACT:**

Epitaxial growth of single crystalline noble metals on dielectric substrates has received tremendous attention recently due to their technological potentials as low loss plasmonic materials. Currently there are two different growth approaches, each with its strengths and weaknesses. One adopts a sophisticated molecular beam epitaxial procedure to grow atomically smooth epitaxial Ag films. However, the procedure is rather slow and becomes impractical to grow films with thickness > 50 nm. Another approach adopts a growth process using rapid e-beam deposition which is capable of growing single crystalline Ag films in the thick regime (> 300 nm). However, the rapid growth procedure makes it difficult to control film thickness precisely, *i.e.*, the method is not applicable to growing thin epitaxial films. Here we report a universal approach to grow atomically smooth epitaxial Ag films with precise thickness control from a few monolayers to the optically thick regime, overcoming the limitations of the two aforementioned methods. In addition, we develop an *in-situ* growth of aluminum oxide as the capping layer which exhibits excellent properties protecting the epitaxial Ag films. The performance of the epitaxial Ag films as a function of the film thickness is investigated by directly measuring the propagation length of the surface plasmon polaritons (SPPs) as well as their device performance to support a waveguide plasmonic nanolaser in infrared incorporating an InGaAsP quantum well as the gain media.

**Text:**

Epitaxial growth of noble metals on semiconductors (or insulators) is fundamentally different from conventional semiconductor/semiconductor or metal/metal epitaxy due to the dramatic surface energy differences between noble metals and semiconductors. The thermodynamic tendency is to form 3D islands to minimize the surface area/volume ratio. This is the reason that even when atomically clean single crystalline semiconductor substrates are used, 3D islands are formed under typical growth conditions.

The first breakthrough came in 1996 when Smith *et al* demonstrated the growth of an epitaxial Ag film on a GaAs substrate using a 2-step procedure involving (a) a low temperature (LT, ~100K) depositing, followed by (b) a room temperature (RT) annealing.[1] The LT deposition quenches the atom diffusion, leading to a high density compact nanocluster film (cluster size ~ 2 nm). The RT annealing allows the Ag atoms to be mobile again, leading to the smoothening of the Ag (111) surface which is a low surface energy surface. Note that in step (b), the formation of large 3D islands is kinetically blocked since it requires moving a large number of atoms over a long distance. By repeating such a 2-step process, one can achieve growth of Ag films of a moderate thickness (25 – 45 nm).[2,3] With proper capping, Lu *et al* used such an atomically smooth Ag film to realize the first continuous wave (CW) operation of low threshold plasmonic nanolasers in 2012.[2] Since in each 2-step cycle, only a few nm (typically 4 or 5 nm) are grown and each cycle takes several hours (due to cooling and warming), such an elaborate procedure becomes totally impractical to grow optically thick film (> 150 nm) which is desirable for many applications.

To grow optically thick films (*e.g.* submicron thick films), an alternative approach has been adopted.[4] In this approach, a very rapid deposition rate (typically > 1 nm/sec using e-beam deposition) is used while the sample is held at a moderately high temperature (e.g. 300 °C).[5, 6] The very high deposition rate can effectively increase the nucleation density, preventing 3D island coarsening. The high sample temperature enhances the atom mobility thus enhancing the smoothening efficiency. This approach is based on a balancing act. The key is achieving a high nucleation density to form a compact film without the formation of large 3D islands. At a fixed deposition rate, if the sample temperature is higher, the nucleation density would be lower and it becomes difficult to form a compact film. Thus, at a higher growth temperature, higher deposition rate is required to maintain the required nucleation density for compact film growth. In this approach, there is likely a minimum thickness (several hundreds of nm) for the formation of compact films. For many advanced applications, precise control of the film thickness from a few to a few hundreds of nm would be necessary.

In order to overcome the limitation of the aforementioned two approaches, here we develop a universal approach, capable to growing epitaxial, atomically smooth Ag films on a Si substrate from a few monolayers to the optically-thick regime by eliminating the repetitive procedures required in the first approach[2, 3, 7] and by increasing the growth rate. This new approach features a moderately high-rate deposition with the sample held at room temperature, followed by a high-temperature-annealing process. An optically-thick (300 nm), atomically smooth, single-crystalline Ag film can be grown in *one cycle*. We apply a whole range of characterization tools including scanning tunneling microscopy (STM), X-ray diffraction (XRD), X-ray photoelectron spectroscopy (XPS), spectroscopic ellipsometry, and a direct measurement of surface plasmon polariton (SPP) propagation distance. Finally, we use such an optically thick Ag film to demonstrate a plasmonic nanolaser at the communication wavelength (1.3 microns) using a metal-insulator-semiconductor (MIS) waveguide structure.

Atomically clean Si (111) surfaces are prepared following the standard procedure (see experimental methods, Part 2.1 of supporting information). A reflection high-energy electron diffraction (RHEED) pattern shown in Figure 1a confirms the well-known 7 × 7 reconstructions on the Si (111) surface. After that, Ag is evaporated onto the Si substrate at RT with a deposition rate of about 3 nm/min, which is about 30 times faster than that adopted by the previous two-step method to grow epitaxial Ag films.[2, 3] As a result, an optically thick (150 nm, or 636 ML) Ag film can be deposited within 1 hour. During the growth, the MBE chamber pressure is kept below $5 \times 10^{-10}$ Torr. The post-growth 150 nm Ag film surface is characterized using RHEED patterns (Figure 1c) and is also compared with that of a 20 ML Ag by the previous two-step method[2, 3] (Figure 1b). The observation of sharp diffraction patterns in these two figures indicates that the long-range single crystalline nature is achieved in both growth methods. Without postgrowth annealing, however, the surface is relatively rough (see inset of Figure 1d), even though sharp streaky patterns are observed with RHEED.

After the growth, the Ag film is transferred to a molybdenum heater equipped with tungsten filaments in the same MBE chamber for *in-situ* annealing. The annealing temperature is kept at 500 ºC for half an hour (the chamber vacuum is kept below $1 \times 10^{-9}$ Torr during the annealing process). The post-growth films are studied by the *in-situ* scanning tunneling microscopy (STM). After annealing, the thick Ag film surface demonstrates atomic-scale surface flatness, as evident from the triangular atomic-layer terraces shown in the STM images for 150 nm (Figure 1d) and 300 nm (Figure 1e) films, respectively. Statistical analysis of surface height distributions is shown in Figure 1f. We also explore the annealing effect at different temperatures and find out that annealing at 500 ºC yields an optimal surface quality according to STM studies (Supporting Information, Figure S1).

To protect the surface of the epitaxial Ag film before taking it to the ambient environment, we develop an *in-situ* growth of a dense aluminum oxide layer. The method entails first epitaxial growth of 7 ML (~1.6 nm) Al on top of Ag, followed by exposure to high purity oxygen under $1.5 \times 10^{-6}$ Torr pressure for 10 minutes. This procedure yields a high-quality, densified $AlO_x$ layer on top of the Ag film. We compare the quality of such *in-situ* growth of aluminum oxide to that of an *ex-situ* oxidized Al layer in the atmosphere using XPS. The Al 2*p* spectra from both samples (150 nm thick Ag (111) films capped by 7 ML Al) are shown in Figure 2. Although the Al 2*p* electron from oxide form ($AlO_x$) is shown at a binding energy of 74.2 eV for both samples, a significant 2*p* electron peak from Al compound mixture (such as $Al(OH)_3$ form) is only observed for the sample oxidized in atmosphere at a lower binding energy of 71.7 eV (Figure 2b). This additional feature indicates the adsorption of $H_2O$ molecules in the atmosphere and the formation of undesired Al compound mixture during the oxidation process, which may increase the density of pin-holes in the capping layer and degrade the underlying Ag film. On the contrary, a pure oxide phase form shown in Figure 2a combined with a higher atom number ratio of oxygen to aluminum (data not shown) suggests that our oxidization method (*in-situ* oxidation by high purity oxygen gas in a vacuum chamber) guarantees adequate oxidization and the formation of a compact capping layer. This property plays a key role to maintain the optical performance of Ag-based plasmonic devices.

After the *in-situ* oxidation of Al, we tested the crystalline quality and the surface roughness of the capped epitaxial thick Ag film by X-ray diffraction (XRD) analysis and atomic force microscopy (AFM), respectively. Figure 3a shows the XRD analysis performed on the 300 nm epitaxial Ag film and only one diffraction peak at 38.2° is observed in the 2*θ* scan (30°−80°), which arises from the Ag (111) crystal plane. The full-width at half maximum (FWHM) of the Ag (111) peak is about 0.2°, suggesting a good crystallinity of the thick Ag film eliminating structural defects including surface roughness and disordered crystal grain boundaries. The capped film retains the atomically smooth surface, as confirmed by the atomic force microscopy (AFM) shown in Figure 3b (for a 150 nm- thick Ag film) and 3c (300 nm thick Ag film). The triangular islands shown in AFM images resembles the terraces observed in the STM image (Figure 1d and 1e) before the *in-situ* AlO$_x$ capping. For comparison, we also deposit both a 150 nm and a 300 nm thick Ag film using a conventional thermal evaporator (Denton, base pressure $2.0 \times 10^{-6}$ Torr, deposition rate ~ 3 nm/min) and their AFM images are shown in Figure 3d and 3e. By comparing the root-mean-square (RMS) roughness of the epitaxial Ag film (300 nm) with that of the polycrystalline one, the former presents a RMS value (0.42 nm, Figure 3c) an order of magnitude smaller than that of the thermal one (4.5 nm, Figure 3e). According to the STM and AFM studies, both the 150 nm and 300 nm films shows a surface morphology as smooth as the one grown by the earlier reported elaborate two-step method.[2, 3]

We also carry out spectroscopic ellipsometry (SE) measurements (supporting information, Part 2.2 and Figure S2). Indeed, the $\varepsilon_2$ data exhibit approximately a factor of 1.5 reductions in the visible frequency range than the widely cited Johnson and Christy data.[8] To confirm the reduced loss, we directly measure the SPP propagation length by measuring the white light interference pattern between two nanogrooves (see schematic and SEM image of the sample in Figure 4a and 4b).[9, 10] Shown in Figure 4c is the scattering spectrum for a double-nanogroove structure fabricated on a 300 nm film with a groove distance of 10 $\mu$m. The inset shows a dark-field image of the double-nanogroove structure. The interference spectrum can be fitted to determine the frequency dependence of the SPP propagation length as shown in Figure 4d. At 632 nm, the SPP propagation length is ~77 $\mu$m and it reaches ~120 $\mu$m at 700 nm. The ultra-long SPP propagation length provides additional evidence for the superior properties of the epitaxially grown thick Ag films. More detailed description and additional measurements for different groove structures can be found in supporting information (Part 2.3 and Figure S3).

To demonstrate the potential of these films in photonic applications, we fabricated a series of metal-insulator-semiconductor (MIS) waveguide nano-lasers operating at the telecom wavelength (~1.3 $\mu$m) based on the epitaxial Ag film. At this wavelength, an optically thick film is required in order to prevent SPP leakage into the substrate. To illustrate this, we systematically vary the Ag film thickness from 25 nm to 300 nm. The MIS structure is constructed by transferring InGaAsP quantum wells (QWs)[11] onto a series of $Al_2O_3$-capped, epitaxially grown Ag films (schematically shown in the inset of Figure 5b). The emission spectra of these MIS structures are examined as shown in Figure 5a. We observe a featureless emission spectrum from the waveguide sitting on top of the 25 nm thick Ag film, indicating that the SPP energy is not well confined within the cavity formed between the waveguide and the 25 nm thick Ag film. In contrast, the emergence of equally spaced SPP Fabry–Pérot (FP) modes with relatively stronger photoluminescence (PL) intensities are observed for a waveguide on top of the 50 nm thick Ag film, indicative of a mitigated propagation loss of SPP FP resonances. Increasing the Ag film thickness further to 150 nm, the PL intensities of SPP FP resonances are significantly enhanced and meanwhile their linewidth becomes narrower than that in the thinner films. The spectral properties of the QW on the 300 nm Ag film are essentially the same as the 150 nm Ag film.

The quality factors (Q-factor) of SPP FP modes can be calculated by $\lambda/\Delta\lambda$ and the values of prominent SPP FP peaks are shown in Figure 5b for Ag films with different thicknesses. The Q-factor is greatly improved for Ag films thicker than 150 nm (Q = 190 for 150 nm film, Q = 166 for 300 nm film) as opposed to thin films (Q = 14 for 25 nm film, Q = 86 for 50 nm film). In Figure 5c, the emission spectra (taken at 4 K) from a waveguide is investigated under quasi-continuous-wave (CW) excitation (785 nm, 1-$\mu$s pulse duration/10-kHz repetition rate) at different power densities. By increasing the pumping density from 0.2 to 0.5 MW/cm$^2$, the SPP FP resonance around 1.3 $\mu$m is boosted by an order of magnitude, accompanied by an observable blue shift and a linewidth narrowing of all resonances. Notably, a significantly enhanced emission intensity is observed near 1.3 $\mu$m when the pumping density is increased to 1.1 MW/cm$^2$, indicating the plasmonic lasing supported by the MIS structure. This lasing behavior is further studied by the S-shaped light-out versus light-in (*L-L*) curve of the emission peak (Figure 5d, log-log scale), confirming a clear transition from spontaneous emission to amplified spontaneous emission.

Since the thickness of Ag film plays a critical role in preventing radiation leakage into the substrate for the MIS structure at telecom wavelength (the SPP skin depth into the metal at telecom wavelength is significantly larger than that in the visible range), we investigate how the radiation energy of the fundamental SPP mode leaks into the Si substrate by plotting the electric field maps of the MIS structure with different Ag film thicknesses. For the MIS structure with a 50 nm Ag film, a considerable amount of radiation energy is observed to leak into the Si substrate (Figure 6a, log scale); for the case of 150 nm Ag film, however, no leakage into the Si substrate is visible (Figure 6b, log scale), suggesting that the 150 nm Ag film is thick enough to prevent radiation leakage into the substrate. This comparison is consistent with the observation that the MIS structure with a 150 nm Ag film presents significantly improved PL emission with respect to that with a 50 nm Ag film shown in Figure 5a.

In order to compare clearly our growth method with previous works to get high quality Ag films, we summarize in Table 1 several main growth conditions (including substrate temperature and deposition rate) and key parameters reflecting the crystalline quality and optical performance of Ag films, such as the measured surface roughness and SPP propagation length. As mentioned above, our rapid MBE growth method features a moderate control of the epitaxial deposition rate and an *in-situ* surface passivation technique which is crucial for the robust performance of Ag nanostructures. As shown in Table 1, the moderate deposition rate (3 nm/min) overcomes the difficulty to obtain optically thick Ag film using traditional tedious MBE approach (0.1 nm/min) while guarantees a precise control of the film thickness hardly achievable by fast sputtering or thermal growth method (~100 nm/min). We used our new procedure to grow atomically smooth films with thickness ranging from 20 nm to 300 nm. To achieve good growth of a few atomic layers epitaxial film, we need to follow the original recipe for low temperature growth followed by room temperature annealing[2, 3]. Moreover, the atomically smooth passivation surface and high crystalline quality of Ag films by the rapid MBE growth method provides superior optical performances with a notably long SPP propagation length (Figure S3).

Table 1. Comparison of different growth methods of Ag film. Note that the SPP propagation length depends on the film thickness when the film thickness is less than 100 nm.

| Methods | Substrate temperature [°C] | Annealing temperature [°C] | Base vacuum (torr) | Growth rate (nm/min) | Thickness (nm) | Surface roughness (RMS, nm) | XRD Ag (111) FWHM (°) | SPP Propagation length @632 nm | Capping method | Refs |
|---|---|---|---|---|---|---|---|---|---|---|
| **MBE 1** | ~−185 | RT | ~$10^{-11}$ | 0.1 | 45 | 0.36 | N/A | ~22 μm | *in-situ* Al | ref[2, 3] |
| **Sputtering 1** | 350 | 350 | ~$10^{-8}$ | ~100 | 97 | 0.82 | 0.3 | ~35 μm | N/A | ref[6] |
| **Sputtering 2** | 300 | 300 | ~$10^{-7}$ | ~90-100 | 100 | 0.3 | 0.116 | ~29 μm@610 nm | N/A | ref[4] |
| **Thermal 1** | 360 | 360 | ~$10^{-6}$ | ~100 | 200 | 0.43 | 0.19 | ~50 μm | N/A | ref[12] |
| **Thermal 2** | 75 | N/A | ~$10^{-5}$ | 0.6-1.2 | 250 | 4.7 | N/A | ~4 μm | N/A | ref[13] |
| **Template stripping** | 75 | N/A | ~$10^{-5}$ | 0.6-1.2 | 200 | 0.36 | N/A | ~35 μm | N/A | ref[13] |
| **Electron-gun** | RT | 340 | ~$10^{-6}$ | 18-30 | 200 | 0.46 | 0.3 | N/A | N/A | ref[14] |
| **Synthesis 1** | N/A | N/A | N/A | N/A | 290 | N/A | N/A | ~11 μm @534 nm | N/A | ref[15] |
| **Synthesis 2** | N/A | N/A | N/A | N/A | 290 | 0.5 | <0.05 | ~80 μm | N/A | ref[10] |
| **MBE 2** | RT | 500 | ~$10^{-11}$ | 3 | 300 | 0.42 | 0.2 | ~80 μm | *in-situ* Al | This work |

In conclusion, we report a new approach to grow atomically smooth, single-crystalline, and optically thick Ag films achieved at a moderately high deposition rate (~3 nm/min). We demonstrate that the capability to achieve optically thick films with good epitaxial quality is critical for achieving the low-threshold, telecom wavelength plasmonic lasing based on a MIS waveguide. We expect that this radically improved production efficiency of single-crystalline, thick Ag films to greatly expand their applications in plasmonic integrated circuits.

ASSOCIATED CONTENT

**Supporting Information**. Height distributions of epitaxial Ag film surface morphologies before and after annealing and STM images of epitaxial Ag films (150 nm) annealed under different temperatures (400ºC, 500ºC and 600ºC) are shown in part 1. Optical characterization of epitaxial Ag films, including measurements of optical constants and SPPs propagation length, are presented in part 2. Experimental methods including the process for treating Si (111) substrate, spectroscopic ellipsometry, white-light interference and numerical simulations are included in part 3.

AUTHOR INFORMATION

**Corresponding Author**

*E-mail: shih@physics.utexas.edu

**Notes**

The authors declare no competing financial interest.

ACKNOWLEDGMENT

This work was partially supported by the Welch Foundation (F-1672 and F-1662) and by the National Science Foundation (NSF-DMR-1306878, NSF-ECCS-1408302 and NSF-EFMA-1542747). S, G., X.Q.L and C.K.S also acknowledge support from NT 3.0 program, Ministry of Education, Taiwan.

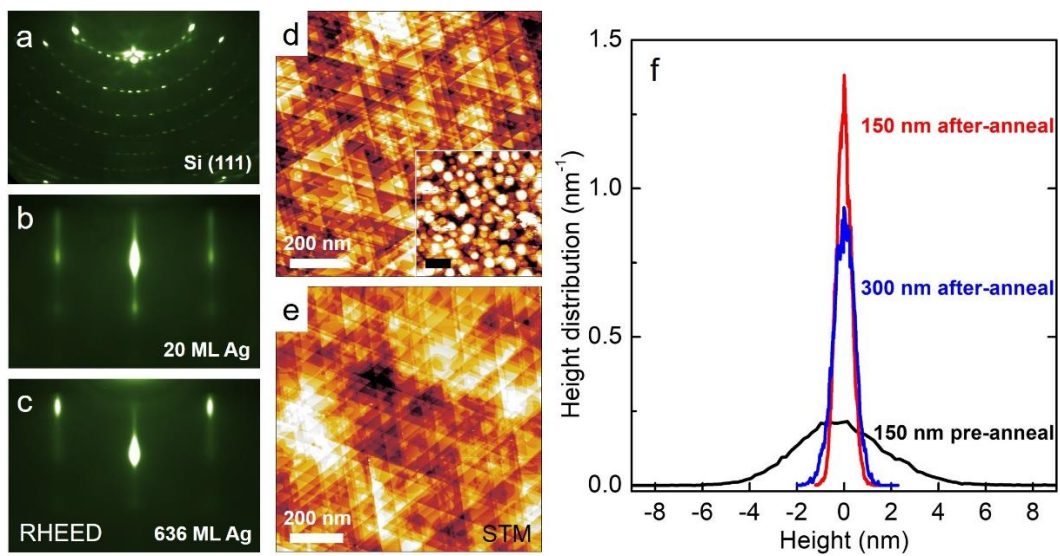

**Figure 1.** (a−c) RHEED patterns of a reconstructed Si (111)-7 × 7 surface, 20 ML Ag (~ 4.7 nm) grown by the reported two-step method and a 636 ML thick Ag film (150 nm) obtained by the new growth method. (d, e) STM images of an epitaxially grown, annealed Ag film (d, 150 nm) and a thicker Ag film (e, 300 nm). The inset of figure d shows a STM scan of the 150 nm Ag film before annealing. Scale bar: 200 nm. (f) Comparison of surface morphologies of epitaxial Ag films before and after annealing (150 nm Ag film before annealing: black; 150 nm Ag film after annealing: red; 300 nm Ag film before annealing: blue). The unit of height distribution function (nm$^{-1}$) is chosen such that the integral of the probability becomes unity.

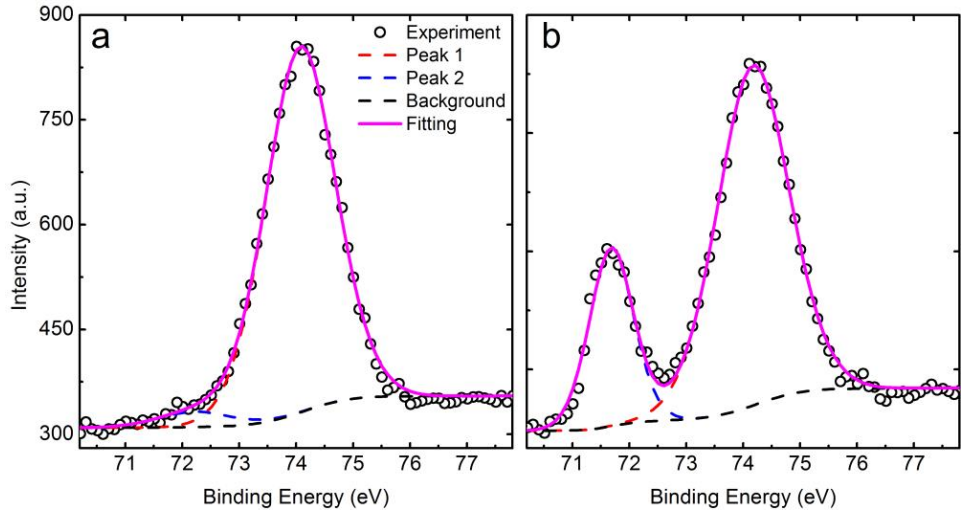

**Figure 2.** X-ray photoelectron spectroscopy (XPS, with Mg Kα source) of the capping layer (AlO$_x$) on top of Ag films oxidized at two different situations: (a) Al-2$p$ electrons from the sample oxidized by high purity oxygen gas in the vacuum chamber. (b) Al-2$p$ electrons from the sample exposed at atmosphere after Al capping. Here the chemical stoichiometry of the AlO$_x$ capping layer is characterized and the binding energies in XPS analysis are standardized by C1s at 284.6 eV.

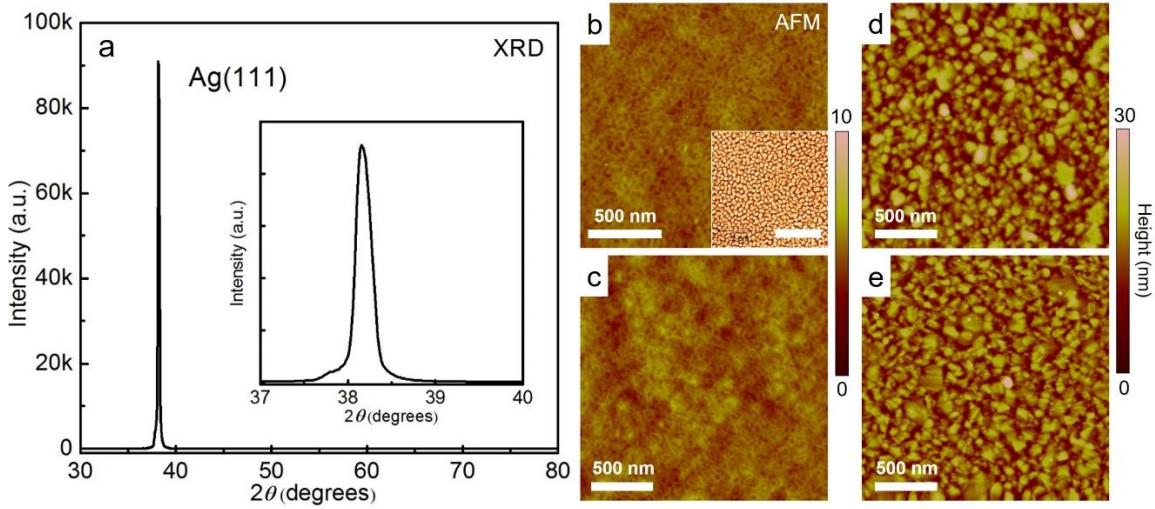

**Figure 3.** (a) XRD 2$\theta$ scan pattern of an epitaxially grown Ag film (300 nm). The inset shows a zoom-in view of the Ag (111) peak. (b, c) AFM images of the 150 nm Ag film and 300 nm Ag film (capping with 7 ML Al and *in-situ* oxidation). The measured root-mean-square (RMS) surface roughness for the 150 nm and 300 nm Ag film is 0.35 nm and 0.42 nm over 2 × 2 $\mu$m$^2$, respectively. The inset of Figure b shows an AFM scan of the 150 nm Ag film before annealing. (d, e) AFM images of thermally deposited polycrystalline Ag films (d: 150 nm, e: 300 nm). The measured RMS for the 150 nm and 300 nm Ag film is 3.6 nm and 4.5 nm over 2 × 2 $\mu$m$^2$, respectively.

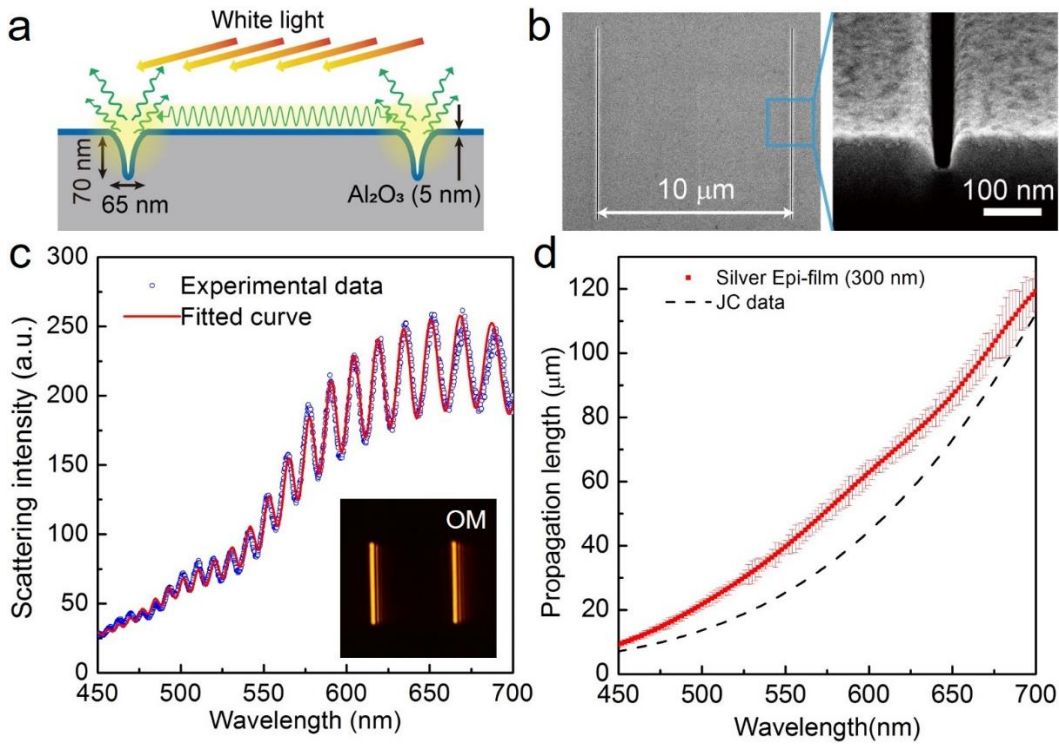

**Figure 4.** (a) Schematic of a WLI setup for measuring the SPP propagation lengths on the epitaxial Ag film (300 nm). The incident angle of the halogen white light is about 75–80º. (b) SEM image of a double-nanogroove structure (top-view) milled by FIB and a zoomed-in image of the cross-section (tilted-view). (c) The scattering spectrum corresponding to the double-nanogroove structure with a distance of 10 $\mu$m shown in figure b. The scattering signal was collected at left hand side of nanogroove by optical microscopy with 100× objective (N.A. = 0.8). The inset shows a dark-field image of a double-nanogroove structure. (d) The SPP propagation lengths on the epitaxial Ag film measured by the WLI method in comparison with those predicted using the JC Ag data. The error bars shown for the WLI method display the s.d. of measurement values from several double-nanogroove structures.

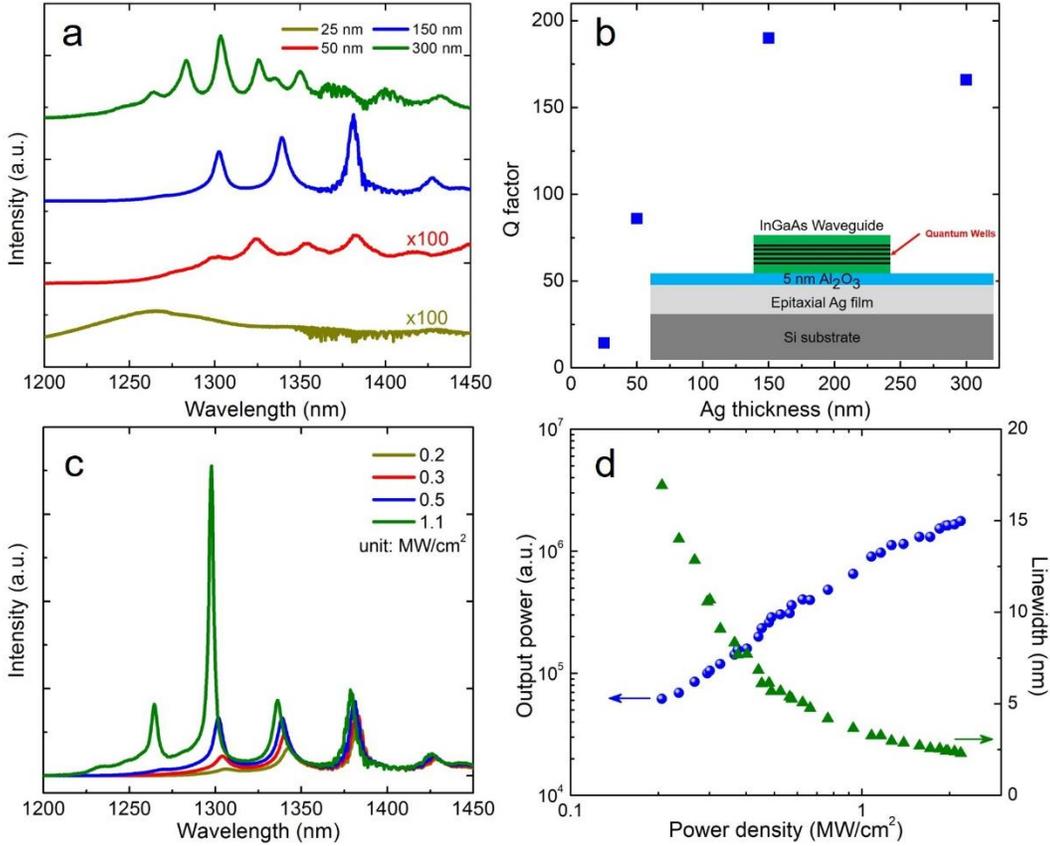

**Figure 5.** (a) Emission spectra of MIS waveguides on Ag platforms with Ag film thickness varying from $t = 25$ to 300 nm. For waveguides on 25 nm and 50 nm Ag films, no lasing is observed. For waveguides on 150 nm and 300 nm Ag films, excitation power density of 0.44 and 0.78 MW/cm$^2$ are used respectively. (b) Q factors calculated by $\lambda/\Delta\lambda$ for Ag films with different thicknesses. The inset shows a schematic of the MIS structure. Top-down fabricated InGaAsP waveguides with embedded quantum wells were transferred onto Al$_2$O$_3$-capped epitaxial Ag platforms. (c) The lasing spectra of a waveguide on the 150 nm Ag film with waveguide width $w= 380$ nm and length $L= 4.8$ $\mu$m. Data are taken at 4 K and at different pumping powers. (d) *L-L* curve (blue circle) and the linewidth narrowing (green triangles) of the lasing peak around 1280 nm shown in c.

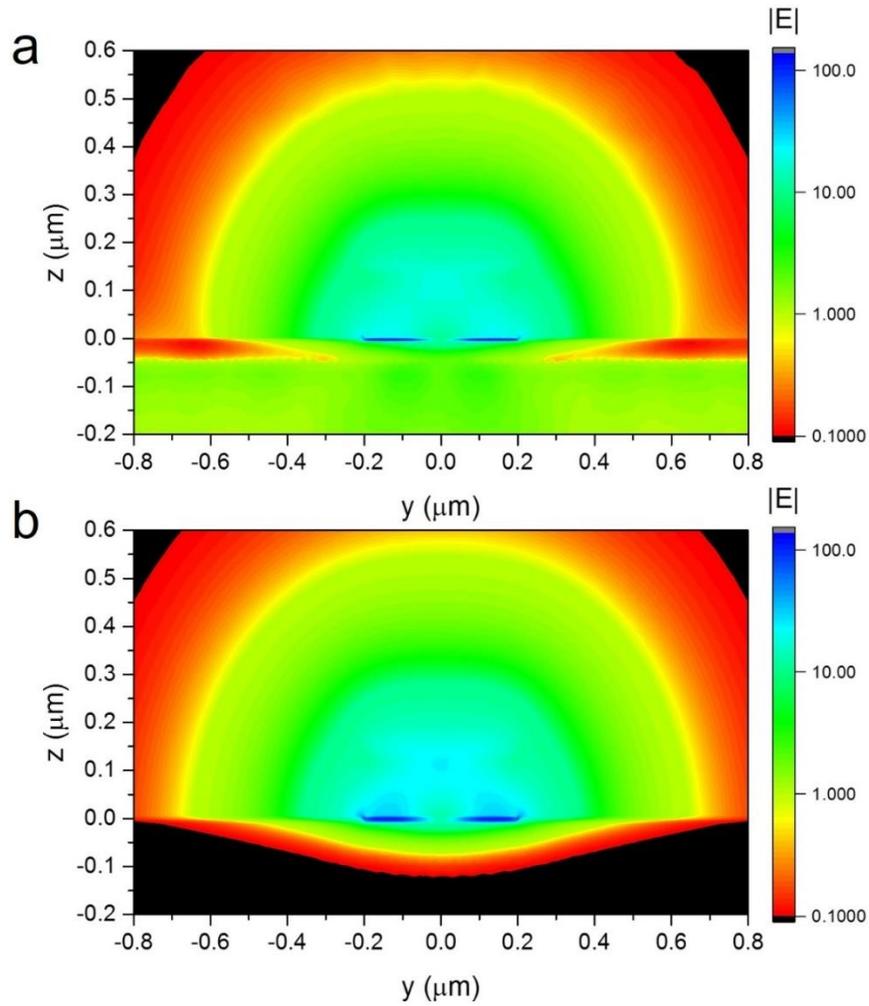

**Figure 6.** (a) Calculated electric field diagram |**E**| showing the radiation leakage of the fundamental mode of the MIS structure shown in Figure 5b with a 50 nm Ag film at $\lambda = 1280$ nm. (b) Calculated electric field diagram |**E**| for the case of 150 nm Ag film. No leakage of radiation energy from SPP mode into the Si substrate is visible. Note here the color legends are shown in log scale.

Supporting Information

# Single Crystalline Silver Films for Plasmonics: From Monolayer to Optically Thick Film

*Fei Cheng,[†] Chien-Ju Lee,[l] Junho Choi,[†] Chun-Yuan Wang,[†,‡] Qiang Zhang,[†] Hui Zhang,[†] Shangjr Gwo,[‡] Wen-Hao Chang,[l] Xiaoqin Li[†] and Chih-Kang Shih[*,†]*

[†]Department of Physics, University of Texas at Austin, Texas 78712 United States

[‡]Department of Physics, National Tsing-Hua University, Hsinchu 30013, Taiwan

[l]Department of Electrophysics, National Chiao Tung University, Hsinchu 30010, Taiwan

[*]E-mail: shih@physics.utexas.edu

**Part 1: Supplementary materials for the morphological characterizations of epitaxial Ag films**

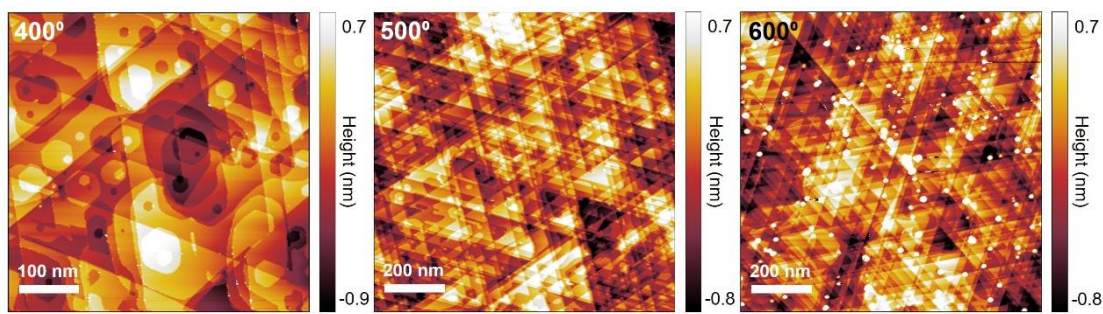

**Figure S1**. STM images of epitaxially grown Ag films (150 nm thick) annealed under different temperatures (left to right: 400°C, 500°C and 600°C). The optimal surface morphology is obtained under the 500°C annealing condition. Under the 400°C annealing condition, we observe a lot of hexagonal Ag islands on the film surface, indicating that the annealing temperature is not high enough. Under the 600°C

annealing condition, we observe instead some small particles on film surface which are the indication of over-annealing.

**Part 2: Supplementary materials for the optical characterization of epitaxial Ag films**

**Part 2.1 The process for treating Si (111) substrate**

We first describe the growth process. A diced arsenic-doped Si (111) wafer (10 × 5 mm$^2$, miscut angle ~0.2º, resistivity = 0.001−0.006 Ω·cm) is used as a substrate accommodated in our home-built UHV-MBE system (base pressure of 5 × 10$^{-11}$ Torr). A dc current (~15 A) is applied on the Si wafer along <110> direction for both outgassing (~2 A) and surface reconstruction purposes. A 2 nm-thick native oxide layer on top of Si is removed during the surface reconstruction process.

**Part 2.2 Spectroscopic ellipsometry (SE) measurement**

In order to explore further the intrinsic property of the thick Ag films, we performed spectroscopic ellipsometry (SE) measurement and analysis on a 300 nm Ag film capped by 7 ML oxidized Al. As shown in Figure S2, the fitted dielectric constants of the thick Ag film are also plotted against the data from a much thinner epitaxial Ag film (45 nm) grown by the elaborate two-step method[1, 2] and the data compiled by Johnson and Christy.[3] The extracted values of $\varepsilon_2$ for the thick Ag film (red solid) are smaller than the JC data (red line-circle) within the whole visible frequency range due to reduced scattering loss from grain boundaries and surface roughness. We note, however, that these values are still slightly larger (15% larger at 550nm) than that of the thin epitaxial Ag film (red dashed) at wavelengths longer than 460 nm, which may be due to a less smooth metal surface of the much thicker Ag film. Shown at the bottom of Figure S3 are the fitting residues (differences between the calculated and experimental effective optical constant) which are small and centered around zero, suggesting a good fit between the chosen model and the measured data.

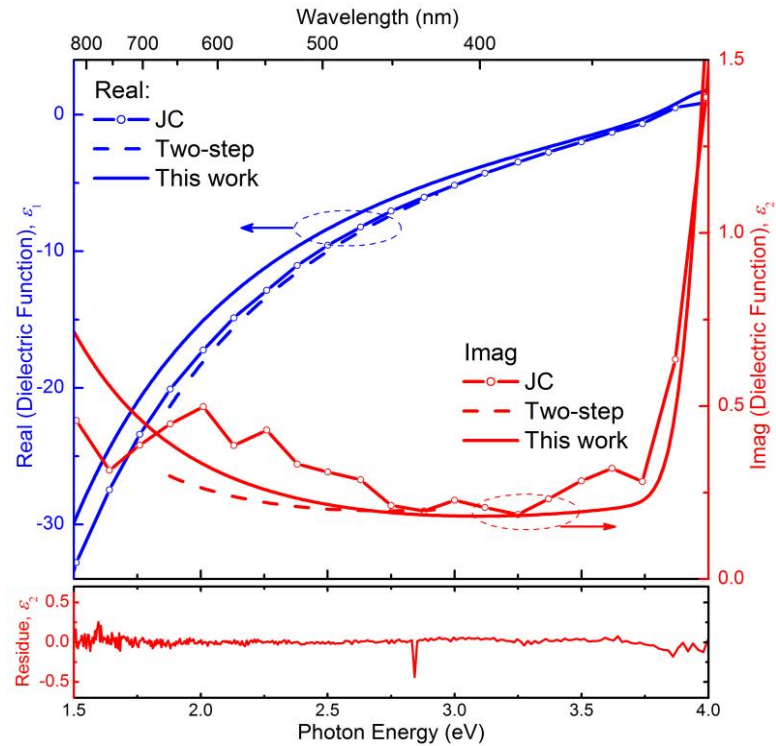

**Figure S2.** Measured dielectric constants of the thick epitaxial Ag film (solid) plotted against those from a 45 nm epitaxial Ag film by the reported two-step method (dash) and values compiled by Johnson and Christy from a 40 nm thermal film (dot-circle). The real part of dielectric constants is represented by blue curves and imaginary part by red curves. The fitting residues of the measured imaginary part for the thick film are also shown at the bottom.

**More details of spectroscopic ellipsometry (SE) measurement**

The SE measurements shown in Figure S2 are performed using a J. A. Woollam M-2000 spectroscopic ellipsometer equipped with a focusing probe attachment with an incident spot size of 300 $\mu$m. Modeling and analysis are performed with the WVASE32 software. We measured the optical constants of a flash-cleaned silicon wafer used in the deposition process as a control sample. Ag films are measured under 3 different incident/collection angles (60º, 65º and 70º) with respect to the normal of the film. For each angle, data are taken at 3 different locations to exclude the location dependence and to verify the spatial homogeneity of the film. To further verify that the Ag film is optically isotropic, we extract the optical constants directly from the raw data before the reverse fitting process. No angular dependence is observed, confirming optical isotropy and no further information is contributed by the multi-angle analysis. The reverse fitting process is performed on data taken at 70º. $\varepsilon_2$ is obtained via fitting of the SE data based on the structural model and analytical equations for dielectric constants, and the values of $\varepsilon_1$ are calculated according to the Kramers–Kronig relations.

**Part 2.3 Measurement of SPP propagation length**

In order to demonstrate the low-loss optical property of the epitaxial Ag film grown using our method, the SPP propagation length in the visible frequency range is measured using the white-light interference (WLI) method,[4] as shown in Figure 4. The measurements are performed on a series of double-nanogroove structures, which are milled by focused ion-beam milling (FIB) on the 300 nm thick Ag film. The SEM images of the double-nanogroove structures and a zoom-in cross-section of a nanogroove are shown in Figure 4b. We would like to mention that the thick epitaxial Ag film allows the fabrication of highly reproducible plasmonic nanostructures using FIB or other methods because grains are absent in such films. For the WLI measurement, a halogen white light excites surface plasmon modes in the nanogrooves with an incident angle of 75–80º, which can be partially coupled out as SPPs propagating along the Ag film and partially coupled to radiation detectable in the far field. The propagating SPPs are reflected back and forth between the nanogrooves and meanwhile couple to the far-field radiation collected by a microscope objective. As shown in Figure 4c, the spectrum exhibits a clear Fabry–Pérot interference pattern as a function of wavelength, which can be fitted by an analytical model from which both the real and imaginary parts of the SPP wavenumber can be retrieved. The propagation length of SPP can then be calculated from the measured SPP wavenumber ($L_{SPP} = 1/2 \times \text{Im}(k_{SPP})$), as shown in Figure 4d. Throughout the full visible spectrum, the measured $L_{SPP}$ values exceed those predicted using the widely cited Johnson and Christy data[3] assuming only ohmic losses. Notably, the $L_{SPP}$ beyond 100 $\mu$m in the red wavelength region is obtained. For actual thermally evaporated polycrystalline films (200 nm), the presence of grain boundaries limits the SPP propagation distance to a few microns in the visible range.[5] As a result, our measurements on the thick epitaxial Ag film exceeds its polycrystalline counterparts by a factor of ~ 10. We also note that our measured $L_{SPP}$ values exceed previously reported values measured on a template-stripped Ag film (200 nm thick) by a factor of 2 (Table 1 in the main text). The data points shown in Figure 4d are averaged measurement values from several double-slit structures, as shown in the following Figure S3.

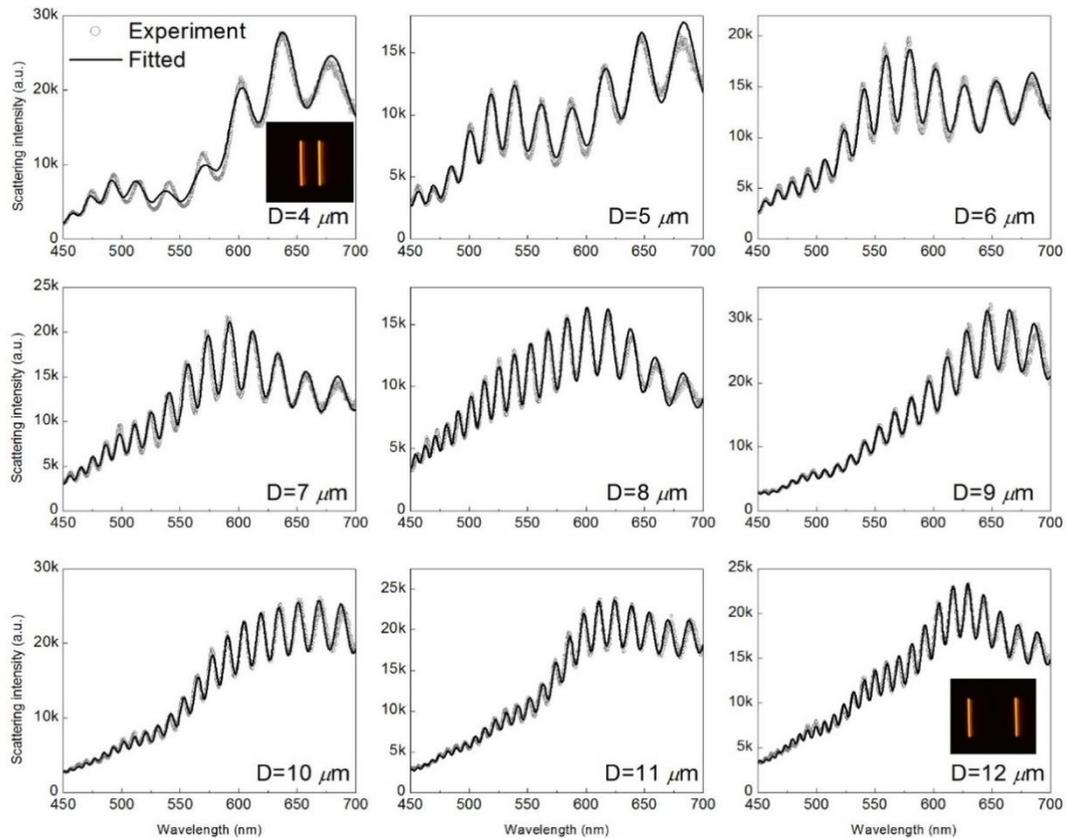

**Figure S3**. Scattering signal of FIB-milled double-nanogroove structures collected by an optical microscopy with 100× objective (N.A. = 0.8). The distances between grooves are indicated in each figure. For the cases of D = 4 μm and 12 μm, the dark field scattering images are shown as insets of the respective figures.